\documentstyle[aps,prb]{revtex}
\begin{document}
\author{V.Ya. Demikhovskii \footnote{demi@phys.unn.runnet.ru} and
D.V. Khomitsky}
\address{Nizhny Novgorod State University \\
 Gagarin Ave. 23, Nizhny Novgorod 603950, Russia}
\title{Quantum states and optics in {\it n}~- and {\it p}~-type
heterojunctions with lateral surface quantum dot (antidot) superlattice
subjected to perpendicular magnetic field}

\maketitle
\begin{abstract}
The studies of quantum states and optics in {\it n}- and {\it p}-type
heterojunctions with lateral surface quantum dot (antidots) superlattice and
in the presence of perpendicular magnetic field are performed. The Azbel'
-- Hofstadter problem is solved for electrons in conduction band and for
holes in valence band described by $4 \times 4$ Luttinger Hamiltonian.
Under the conditions of non-interactive Landau levels the set of magnetic
subbands is obtained for separate electron and hole levels in wide interval
of magnetic field. The influence of spin-orbit interaction onto wavefunctions
and energy spectrum in hole magnetic subbands has been investigated.
The probabilities of transitions between quantum states in magnetic subbands
and monolayer of impurities located inside heterojunction are calculated
for two cases: transitions between electron states and acceptors and between
hole states and donors. The set of parameters (superlattice periods, amplitude
of periodic potential and magntitude of magnetic field, etc.) required for
experimental observation of magnetic subbands is found.
\end{abstract}

\vspace{0.5cm}
PACS number(s): \ 73.21.-b, 73.21 Cd, 78.67.-n

\vspace{0.5cm}

\section{Introduction}

The problem of quantum states of 2D Bloch electrons subjected into magnetic
field remains actual over several last decades. The fascinating physical
phenomena occuring here are caused by interaction of lattice periodic
potential which leads to band structure of spectrum with vector potential of
uniform magnetic field tending to form discrete energy levels. The crucial
parameter determining the nature of quantum states in this problem is
a magnetic flux through lattice elementary cell. If this flux equals rational
number $p/q$ of flux quanta $\Phi_0=2\pi\hbar c/|e|$ ($p$ and $q$ are mutual
prime integers), the magnetic translations form a group. In this case it
becomes possible to determine the transformational law for wavefunction under
translations, the magnetic Brillouin zone and the subband energy spectrum.
When the amplitude of periodic potential $V_0$ is smaller then cyclotron
energy $\hbar \omega _c$ one can neglect the influence of neighbouring Landau
levels and may obtain the set of magnetic subbands arising from single level
\cite{Thouless}. In this scheme every Landau level splits into $p$ subbands
with degeneracy degree $q$. If it becomes needful to include the interaction
between Landau levels, numerical methods are usually applied for calculation
of quantum states \cite{Silb,Geisel,DP}.

During last years several significant theoretical aspects of discussed problem
have been investigated. In particular, quantization of Hall conductance in
the presence of additional periodic potential has been studied in
\cite{Thouless,Geisel,Usov}. One might expect that each of magnetic subbands
to give a Hall conductance equal to $e^2/ph$, but according to Laughlin each
subband must carry an integer multiple of the Hall current carried by the
entire Landau level. The analytical approaches applied in
\cite{Wiegman,Hatsugai,Krasovsky} have indicated that the problem of Bloch
quantum states in magnetic field can be studied using the methods of
Bethe - Ansatz. In particular, the "mid" band spectrum of the model and
the Bloch wavefunction can be found analytically from Bethe - Ansatz equation
that is typical for completely integrable quantum systems.
In more complicated models describing Bloch electrons in magnetic field
the manifestation of quantum chaos has been discovered
\cite{Eroms,Petscel,Ketzmerick}.

Recently the number of experimental studies have been performed in order
to investigate electron quantum states in 2D heterojunctions with lateral
surface superlattice of quantum dots (abtidots). Such a system is convenient
for investigation of both classical effects (commensurability of lattice
periods and cyclotron radius, transition to chaos, etc.) and of energy
spectrum consisting of magnetic subbands. For example, in \cite{Eroms,Weiss}
the oscillations of longitude magnetoresistance have been detected under
the conditions where classical cyclotron radius $2R_c$ envelopes the integer
number of antidots or numerous reflections from one antidot occur. The first
experimental evidences of Landau levels splitted into the set of magnetic
subbands have been obtained in \cite{Schl} by longitude magnetoresistance
studies. Then, the measurements of Hall resistance in subband energy spectrum
have also been performed and reportered this year \cite{Albrecht}.

Besides the magnetoresistance measurements, the attempts of magnetooptical
studies of interband transitions between conduction band and acceptor
impurities \cite{Kukushkin} have been performed in $n$ -type
heterostructures. The experiments in $p$-type heterojunctions without periodic
potential have also become possible due to the progress in technology which
substantially improved the quality of the $p$ channels in GaAs/AlGaAs
heterojunctions \cite{Volkov}. Thus, almost all intriguing
phenomena found for 2D electron system were also observed in 2D hole channels.
The specific traits of hole quantum states which have caused the interest to
them may be briefly described as non-trivial effects of symmetry and
spin-orbit interaction. It is known that in the absence of magnetic field
the electron spectrum in symmetrical quantum well is twofold degenerate with
respect to spin. On the opposite, in asymmetrical heterojunction grown, for
example, in $z$ direction where $V(z)\ne V(-z)$ the relativistic orbital
interaction of electron magnetic moment and macroscopic heterojunction
potential leads to the breakdown of spin degeneracy. Only twofold Kramers
degeneracy
$E\left({\bf k},\uparrow \right)=E\left({\bf -k},\downarrow \right)$ remains.

In order to obtain transparent and valuable results from transport and optical
experiments, one may need to choose the set of parameters (superlattice
periods, value of magnetic field and amplitude of periodic potential, etc.)
which provide a sharp, easily  distinguishable picture of non-overlapped
magnetic subbands originating from particular Landau level. Such energy
spectra and wavefunctions are studied in the present paper together with
calculation of matrix elements for interband transitions between magnetic
subbands and impurities. In Sec. II we obtain electronic spectra and
wavefunctions for spinless electrons in $n$-type heterojunction with
lateral superlattice of quantum antidots and perpendicular magnetic field.
In this case we assume a simple parabolic electron spectrum and here it is
possible to neglect spin-orbit coupling. Sec. III is devoted to the studies of
hole quantum states in $p$-type heterojunction subjected to magnetic field
(Subsec. IIIA) and both to magnetic field and periodic potential of quantum
dots superlattice (Subsec. IIIB). On the opposite to the case of electron
states, the spin-orbit coupling is included here which is principal for
description of holes in semiconductors. We calculate hole magnetic subbands
and wavefunctions assuming several levels of size quantization in
heterojunction. Then in Sec. IV we calculate the matrix elements for
transitions between electron magnetic subbands and monolayer of acceptors
located inside heterojunction, and for transitions between hole magnetic
subbands and monolayer of donors. The huge difference in magnitude of matrix
elements corresponding to different magnetic subbands was found and
the dependence of transition probabilities on polarization was investigated.
These results may be used for identification of complicated magnetic subband
spectra in magnetooptical experiments. The summary of our results is given
in Sec. V.

\section{Electron states in heterojunction with lateral superlattice and
perpendicular magnetic field}

The Bloch electron wavefunction at the $\Gamma$-point of conduction band
of {\it n}-type GaAs/AlGaAs heterojunction can be constructed in
envelope function approximation as a product of $s$ -\ type atomic function
$s({\bf r})$ and envelope function $\psi_{k_xk_y}$:
\begin{equation}
\label{blochtot}
\Psi_{k_xk_y}({\bf r})=\psi_{k_xk_y}s({\bf r}).
\end{equation}

\noindent In Eq.(\ref{blochtot}) $\psi_{k_xk_y}$ stands for envelope function
which satisfies to Schr\" odinger equation for spinless electron with
effective mass $m_{eff}$ subjected into uniform magnetic field ${\bf H}||Oz$
and periodic potential of lateral surface superlattice which can be chosen in
the form \cite{DP}
\begin{equation}
\label{vxy}
V(x,y)=V_0\cos^2\frac{\pi x}{a}\cos^2\frac{\pi y}{a}.
\end{equation}

\noindent Here $a$ is superlattice period and the case $V_0<0\ (>0)$
corresponds to periodic electric potential generated by quantum dots
(antidots) superlattice. The Bloch electron wavefunction $\psi_{k_xk_y}$
which is the solution of Schr\"odinger equation with vector potential
\begin{equation}
\label{gauge}
{\bf A}=(0,Hx,0)
\end{equation}

\noindent and periodic potential (\ref{vxy}) satisfies to
the generalized Bloch boundary conditions (Peierls conditions)
\begin{eqnarray}
\nonumber
\psi_{k_x k_y}(x,y,z)=\psi_{k_x k_y}(x+qa,y+a,z)\exp(-ik_x qa)\times \\
\label{pei}
\times \exp(-ik_y a)\exp(-2\pi ipy/a),
\end{eqnarray}

\noindent In Eq.(\ref{pei}) the magnetic flux $\Phi=Ha^2$ through superlattice
elementary cell equals rational number $p/q$ of flux quanta
$\Phi_0=2\pi \hbar c/e$ where $p$ and $q$ are mutually prime integers.
Following \cite{DP}, we write the electron wavefunction as
\begin{eqnarray}
\nonumber
\psi_{k_x k_y}({\bf r})=\frac{1}{La\sqrt{q}}
\sum_{N=0}^{\infty}\sum_{n=1}^p C_{Nn}(k_x,k_y)
\sum_{l=-L/2}^{L/2}u_N\left(\frac{x-x_0-lqa-nqa/p}{\ell_H} \right)\times \\
\label{psiel}
\times \exp\left(ik_x\left[lqa+\frac{nqa}{p}\right]\right)
\exp\left(2\pi iy\frac{lp+n}{a}\right)\exp(ik_y y)
\end{eqnarray}

\noindent where $\ell_H=\sqrt{\hbar c/eH}$ is a magnetic length,
$x_0=k_y \ell_H^2$ and $u_N(x)$ is $N$-th harmonic oscillator wavefunction.
We substitute wavefunction (5) into Scr\"odinger equation and after standard
quantum-mechanical projection onto basis in Hilbert space the eigenvalue
problem for coefficients $C_{Nn}$ is obtained. The energy spectrum
$\varepsilon_{Nn}(k_x,k_y)$ forms a set of $p$ magnetic subbands
($n=1,\ldots,p$) for each Landau level $N$ \cite{Thouless}. Electron spectrum
at $k_x=k_y=0$ in antodot lattice with $a=80 \ nm$ and $V_0=20 \ meV$ is shown
on Fig.1 for three Landau levels: $N=0$, $N=1$, and for $N=4$ on the inset.
At high magnetic fields when $\hbar\omega_c>V_0$ these subbands are very
narrow and look like a set of almost discrete levels where the energy is
practically independent with respect to $k_x$ and $k_y$. After comparing
the structure of subbands originating from different levels it is clearly seen
that the internal structure of splitted Landau level varies with the respect
to level number $N$, namely, the total level splitting decreases and
the position of clustering point moves to the higher energies.

It should be stressed that the spectrum on Fig.1 is obtained for periodic
potential (\ref{vxy}) which sign is a constant defined by $V_0$. Thus, our
spectrum differs from those for periodic potential of the form
$V(x,y)=V_0\left(\cos 2\pi x/a + \cos 2\pi y/a \right)$ where for energy
dependence with respect to $q/p$ one gets a Hofstadter butterfly scaled by
Laguerre polynomial $L_N$ \cite{Thouless}. Considering the more realistic
potential of quantum dots (antidots) (\ref{vxy}), one obtains the energy
spectrum (versus $q/p$ ) which significantly differs from Hofstadter
butterfly. When the condition $V_0<\hbar \omega _c$ is satisfied, we can
use the approximation of non-interactive Landau levels and thus study
the splitting of each level independently, neglecting the summation over
$N$ in (\ref{psiel}). The corresponding matrix equation for coefficients
$C_{n}(k_x,k_y)$ can be written as
\begin{eqnarray}
\nonumber
\frac 12 e^{ik_xa \frac qp}\Biggl[e^{-\frac{\pi q}{2p}}
L_N \left(\pi \frac qp\right)+
\cos\left(2\pi \frac qp\left[n+\frac 12\right]+k_ya \frac qp \right)
e^{-\pi \frac qp}L_N \left(2\pi \frac qp\right)\Biggr]C_{n+1}+ \\
\nonumber
\frac 12 e^{-ik_xa \frac qp}\Biggl[e^{-\frac{\pi q}{2p}}
L_N \left(\pi \frac qp\right)+
\cos\left(2\pi \frac qp\left[n-\frac 12\right]+k_ya \frac qp \right)
e^{-\pi \frac qp}L_N \left(2\pi \frac qp\right)\Biggr]C_{n-1}+ \\
\label{silev}
+e^{-\frac{\pi q}{2p}}L_N\left(\pi \frac qp\right)
\cos\left(2\pi n \frac qp +k_ya \frac qp \right)C_n=\varepsilon C_n
\end{eqnarray}

\noindent where cyclic boundary condition $C_{n+p}=C_n$ is assumed and
cyclotron energy $\hbar \omega _c$ is excluded. The spectrum of system
(\ref{silev}) is shown on Fig.2 for three lowest Landau levels $N=0,1,2$.
The representation with respect to reciprocal number of flux quanta $q/p$
provides the information on energy spectrum in wider interval of magnetic
fields compared with Fig.1 and thus allows us to visualize the energy spectrum
both at low ($q/p \approx 1$, right side of Fig.2) and at high ($q/p \ll 1$,
left side of Fig.2) magnetic fields. The spectrum on Fig.2 indicates that at
intermediate magnetic fields $0.1<q/p<1$ the splitting of Landau levels varies
due to non-monotonous behaviour of Laguerre polynomials. For small values of
$p$ and $q$ (for example, $q/p=1/4,1/3,1/2,2/3,\ldots$) the spectrum consists
of $p$ relatively wide non-overlapping magnetic subbands while for $q/p \ll 1$
(high magnetic fields) the energy subbands are very narrow in accordance with
those shown on Fig.1.

Considering our further studies of magnetooptical transitions from acceptors
to conduction band, we should be aware of electron wavefunction behaviour
in a single superlattice cell compared with those for acceptor. The typical
Bohr radius of shallow acceptor in GaAs $r_A \approx 3 nm$ which is much
smaller then superlattice period $a=80 nm$. This leads to strong dependence
of matrix elements on particular position of acceptor inside the superlattice
cell. On Fig.3 we show the wavefunctions for two magnetic subbands marked by
arrows on Fig.1 which originate from the lowest Landau level. Fig.3a
corresponds to the 4th subband located in the region of subbands clustering
and Fig.3b is plotted for the 20th (highest) subband. For simplicity we show
only the positive values of real part of wavefunction. Hereafter the darker
areas on contourplots are related to greater values of wavefunctions.
The circle on Fig.3a illustrates relative scale of acceptor and electron
wavefunctions. One can clearly see that their overlapping crucially depends
on the position of acceptor and the influence of this overlapping on matrix
elements will be studied in Sec. IV.

\section{Hole quantum states in the presence of lateral superlattice and
magnetic field}

\subsection{Hole Landau quantum states in $p$-~type heterojunction without
periodic potential}

We now consider the upper fourfold edge of GaAs
$p$-~like valence band at ${\bf k}=0$. Its bulk band structure in the presence
of a magnetic field applied in $\langle 001 \rangle$ direction (hereafter
denoted by $z$) is described in axial approximation in terms of
$4\times 4$ effective Luttinger Hamiltonian \cite{Luttinger,Broido}
\begin{eqnarray}
\label{lutt}
\vspace{1 cm}
H_L= \left[ \matrix{
H_{11} & {\overline \gamma}\sqrt{3}(eH/c)a^2 & \gamma_3\sqrt{6eH/c}\ k_z a
& 0 \cr
\ & H_{22} & 0 & -\gamma_3\sqrt{6eH/c}\ k_z a \cr
\ & \ & H_{33} & {\overline \gamma}\sqrt{3}(eH/c)a^2 \cr
\ & \ & \ & H_{44}
} \right],
\end{eqnarray}

\noindent where

\begin{eqnarray}
\nonumber
H_{11}=-(\gamma_1/2-\gamma_2)k_z^2-(eH/c)\left[(\gamma_1+\gamma_2)
\left(a^+ a + \frac{1}{2} \right)+\frac{3}{2}\kappa \right], \\
\nonumber
H_{22}=-(\gamma_1/2+\gamma_2)k_z^2-(eH/c)\left[(\gamma_1-\gamma_2)
\left(a^+ a + \frac{1}{2} \right)-\frac{1}{2}\kappa \right], \\
\nonumber
H_{33}=-(\gamma_1/2+\gamma_2)k_z^2-(eH/c)\left[(\gamma_1-\gamma_2)
\left(a^+ a + \frac{1}{2} \right)+\frac{1}{2}\kappa \right], \\
\nonumber
H_{44}=-(\gamma_1/2-\gamma_2)k_z^2-(eH/c)\left[(\gamma_1+\gamma_2)
\left(a^+ a + \frac{1}{2} \right)-\frac{3}{2}\kappa \right],
\end{eqnarray}

\noindent and the lower half of the matrix is obtained by Hermitian
conjugation. Here atomic units $\hbar=m_0=1$ are used and the hole energy is
measured as negative, $e$ is a module of elementary
charge, ${\overline \gamma}=(\gamma_2+\gamma_3)/2$, \ $H$ stands for
magnitude of magnetic field, $a^+$ and $a$ are harmonic oscillator raising
and lowering operators. The band parameters appearing in matrix (\ref{lutt})
are taken from \cite{Broido}: \ $\gamma_1=6.85$, $\gamma_2=2.1$,
$\gamma_3=2.9$, and  $\kappa=1.2$. The Luttinger Hamiltonian (\ref{lutt})
is written in a basis of $p$-~like atomic functions $v_j({\bf r})$ which
transform as a set of eigenfunctions for angular momentum operator $J=3/2$.
These $|J;m_J\rangle$ basis fuctions may be written as following:
\vspace{0.5cm}
\begin{equation}
\label{jmj}
\cases{
v_1=\mid\frac32;\frac32\rangle=\left|-\sqrt{1/2}(x+iy)\uparrow
\right\rangle, &  \cr
v_2=\mid\frac32;-\frac12\rangle=\left|-\sqrt{1/6}(x-iy)\uparrow
-\sqrt{2/3}z\downarrow\right\rangle, &  \cr
v_3=\mid\frac32;\frac12\rangle=\left|\sqrt{1/6}(x+iy)\downarrow
-\sqrt{2/3}z\uparrow\right\rangle, & \cr
v_4=\mid\frac32;-\frac32\rangle=\left|-\sqrt{1/2}(x-iy)\downarrow
\right\rangle, & \cr }
\end{equation}
\vspace{0.5cm}

\noindent where the arrows indicate $z$-projection of spin.

The holes in GaAs/AlGaAs $p$-~type heterojunction grown in $z$ direction which
is parallel to the magnetic field are confined by potential $V_h(z)$ which is
a smoothly varying function with triangular shape. It should be noted that
such a shape does not have inversion symmetry, i.e. $V_h(z)\ne V_h(-z)$ which
leads to the breakdown of twofold spin degeneracy and to the splitting of
energy levels of effective Hamiltonian
\begin{equation}
\label{heff}
H_{eff}=H_L(a^+,a,k_z)+V_h(z)
\end{equation}

\noindent even at the absence of magnetic field \cite{Broido}. The lack of
inversion symmetry of the atomic potential of GaAs crystal lattice is present
also in bulk material and is described by linear $k$-terms in Luttinger
Hamiltonian. However, the effects caused by these terms (the displacement of
subbands maxima in {\bf k}-space \cite{Rashba,Sham}) are neglegible compared
with those induced by heterostructure potential and thus are not considered
here.

The solution of the effective-mass equation with Hamiltonian
(\ref{heff}) in each of two materials constituting the heterojunction
may be written as a four-componenet vector of envelope functions in
$|J;m_J\rangle$ basis (\ref{jmj}). As it was shown by Luttinger
\cite{Luttinger}, in the presence of magnetic field and under axial
approximation one can distinguish the eigenstates of operator (\ref{heff}) by
discrete quantum number $n$ which defines the particular set of Landau quantum
states. These states have $k_y$-component of momentum under Landau gauge
(\ref{gauge}) and in the presence of heterostructure potential
the $k_z$-component is replaced by operator $k_z=-i\partial / \partial z$.
Hence, the eigenstate $F_{nk_y}$ of operator (\ref{heff}) consists of four
envelope functions $c_j(z)$, $j=1,2,3,4$ \cite{Broido} and the hole
wavefunction is written as
\begin{equation}
\nonumber
\Psi_{nk_y}=\sum_{j=1}^4 F_{jnk_y}v_j
\end{equation}

\noindent where $v_j$ is a $|J;m_J\rangle$ basis function. Here one can write
\begin{equation}
\label{fn}
F_{nk_y}=\left(c_1(z)\phi_{n-2,k_y}, \ c_2(z)\phi_{n,k_y},\ c_3(z)
\phi_{n-1,k_y},\ c_4(z)\phi_{n+1,k_y} \right).
\end{equation}

\noindent In Eq.(\ref{fn}) $\phi_{nk_y}(x,y)=e^{ik_y}u_n(x)$ is
Landau quantum state, and envelope functions $c_j(z)$ vanishe for negative
indexes $n$. For example, for $n=-1$ one can obtain $F_{-1}=(0,0,0,c_4(z)
\phi_0)$, for $n=0$ the solution $F_0=(0,c_2(z)\phi_0,0,c_4(z)\phi_1)$, and
for $n \ge 2$ all four components of (\ref{fn}) will be nonzero. It should
be noted that the particular classification of solutions $F_n$ may be chosen
in a different way \cite{Volkov} which leads to changes in notation only.

We first observe that for $H=0$ the Hamiltonian (\ref{heff}) becomes diagonal
with elements
\begin{eqnarray}
\nonumber
H_h=-(\gamma_1/2-\gamma_2)\frac{d^2}{dz^2}+V_h(z), \\
\nonumber
H_l=-(\gamma_1/2+\gamma_2)\frac{d^2}{dz^2}+V_h(z)
\end{eqnarray}

\noindent that yields an infinite set of doubly degenerate heavy and light
hole subband energies and eigenfunctions $c_{\nu j}(z),\nu =1,2,\ldots $
These functions are usually obtained by solving Schr\"odinger and Poisson
equations self-consistently. As a result, the shape of potential $V(z)$ has
a varying gradient which reflects the changes in electric field
inside the heterojunction \cite{Volkov,Broido}. Thus, the precise shape of
functions $c_{\nu j}(z)$ differs from the one for the case of uniform electric
field. However, the investigations of energy spectrum and matrix elements
of transitions between 2D Bloch quantum states and impurities require only
the information on overlapping between different localized functions
$c_{\nu j}(z)$, and between them and well-known wavefunctions of impurities.
The intervals of localization for $c_{\nu j}(z)$ can be obtained with high
accuracy for all subbands of size quantization considered in this paper since
the shape of $V(z)$ in single GaAs/AlGaAs heterojunction is well-known and
was taken by us from \cite{Broido}.

For non-zero magnetic field we have a fan chart of Landau levels
originating from each level of size quantization and therefore it is possible
to construct the envelope functions for finite $H$ as
\begin{equation}
\label{fnb}
F_{jk_y}=\sum_{\nu_j n_j}C_{j\nu_j n_j}\phi_{n_jk_y}c_{j\nu_j}.
\end{equation}

\noindent After substituting the function (\ref{fnb}) into Schr\"odinger
equation with Hamiltonian (\ref{heff}) one obtaines an algebraic eigenvalue
problem for coefficients $C_{j\nu_j n_j}$. We restrict ourself to the first
three levels of size quantization which corresponds to consideration of two
heavy- and one light-hole levels. This approximation seems to be valid in
heterojunctions with typical hole concentration $n=5 \times 10^{11} cm^{-2}$
and depletion-layer density $N_{dep}=10^{15} cm^{-3}$
where only the lowest hole level is occupied \cite{Broido,Volkov}. For
each level of size quantization we take into account several Landau levels
shown on Fig.4. Here one can see the electron-like behaviour of light-hole
Landau levels at low magnetic field caused by proximity of second heavy-hole
subband. We assume that the introduction of periodic potential
with amplitude $V_0$ (see the following Subsec.) does not change
$c_{\nu j}(z)$ significantly since $|V_0|$ considered in our paper is much
smaller then size quantization energies. Hence, in our further studies we
use matrix elements of effective Hamiltonian (\ref{heff}) calculated for
the functions $c_{\nu j}(z)$ and size quantization energies from
\cite{Broido}.

\subsection{Bloch quantum states in the presence of lateral surface
superlattice}

The problem of hole quantum states in a $p$-~type heterojunction subjected
into magnetic field and affected by lateral superlattice is described by
Scr\"odinger equation with vector potential (\ref{gauge}) and periodic
potential of lateral superlattice given by (\ref{vxy}). The Hamiltonian of
this problem is a sum of (\ref{heff}) and (\ref{vxy}):
\begin{equation}
\label{htot}
H=H_{eff}+V(x,y)\cdot E,
\end{equation}

\noindent where $E$ being a unit $4\times 4$ \ -~matrix. The eigenvectors of
operator (\ref{htot}) are envelope functions written in $|J;m_J\rangle$
basis (\ref{jmj}). The crucial statement here is the following: as long as
periodoc potential (\ref{vxy}) is applied, every hole envelope function
becomes a Bloch function (in the presence of magnetic field) in $(xy)$ plane
and is classified by $k_x$ and $k_y$ quantum numbers. Hence, one can write
the eigenvector $\Psi^{envelope}_{k_x k_y}({\bf r})$ of operator (\ref{htot})
as
\begin{equation}
\label{psienv}
\Psi^{envelope}_{k_x k_y}({\bf r})=\left(\psi^{(1)}_{k_x k_y}({\bf r}),\
\psi^{(2)}_{k_x k_y}({\bf r}),\ \psi^{(3)}_{k_x k_y}({\bf r}),\
\psi^{(4)}_{k_x k_y}({\bf r})  \right),
\end{equation}

\noindent and the four-component hole wavefunction is
\begin{eqnarray}
\nonumber
\Psi_{k_x,k_y}({\bf r})=
\psi^{(1)}_{k_x k_y}({\bf r})\left|\frac32;\frac32\right\rangle+
\psi^{(2)}_{k_x k_y}({\bf r})\left|\frac32;-\frac12\right\rangle+ \\
\label{psihole}
\psi^{(3)}_{k_x k_y}({\bf r})\left|\frac32;\frac12\right\rangle+
\psi^{(4)}_{k_x k_y}({\bf r})\left|\frac32;-\frac32\right\rangle.
\end{eqnarray}

\noindent It should be mentioned that the translational properties of
each component of
envelope function (\ref{psienv}) are the same as for electron wavefunction
(\ref{psiel}). In particular, (\ref{psienv}) satisfies to Peierls condition
(\ref{pei}). Hence, the hole envelope function may be written in the form
\begin{eqnarray}
\nonumber
\psi^{(j)}_{k_x k_y}({\bf r})=\frac{1}{La\sqrt{q}}
\sum_{S_j}c_{S_j}(z)\sum_{N_j}\sum_{n=1}^p G_{j S_j N_j n}(k_x,k_y)
\sum_{l=-L/2}^{L/2}u_{Nj}\left(\frac{x-x_0-lqa-nqa/p}{\ell_H} \right)\times
\\
\label{psiho}
\times \exp\left(ik_x\left[lqa+\frac{nqa}{p}\right]\right)
\exp\left(2\pi iy\frac{lp+n}{a}\right)\exp(ik_y y),
\end{eqnarray}

\noindent where for particular $|J;m_J\rangle$ projection $j$ we summerize
over size quantization levels $S_j$, over Landau levels $N_j$ and over
magnetic subbands $n$. Then, analogous to the electron problem desribed in
Sec. II, after substituting the wavefunction (\ref{psihole}) into
Schr\"odinger equation with Hamiltonian (\ref{htot}) one obtains
the eigenvalue problem for coefficients $G_{j S_j N_j n}(k_x,k_y)$ and hole
magnetic subbands $\varepsilon_{j S_j N_j n}(k_x,k_y)$:
\begin{equation}
\label{matrix}
\sum_{j'S_j'N_j'n'}\left(
H^{j'S_j'N_j'n'}_{j S_j N_j n}+V_{j S_j N_j n}^{j'S_j'N_j'n'}(p/q,k_x,k_y)
\right)G_{j'S_j'N_j'n'}=\varepsilon G_{j S_j N_j n}.
\end{equation}

\noindent Here the notation $H^{j'S_j'N_j'n'}_{j S_j N_j n}$ is used for
projection of Hamiltonian (\ref{heff}) onto our basis $(j\ S_j\ N_j\ n)$ and
$V_{j S_j N_j n}^{j'S_j'N_j'n'}(p/q,k_x,k_y)$ stands for matrix elements
of periodic potential (\ref{vxy}) calculated in this basis. The spectrum
of system (\ref{matrix}) at the center of magnetic Brilloin zone $k_x=k_y=0$
is shown of Fig.5 for the case of non-overlapped subbands related to
the highest hole levels $n=2+$ and $n=-1-$. Here the sign $+(-)$ refers to
the spin projection of dominating component of $|J;m_J\rangle$  basis
\cite{Broido,Volkov}. Similar to the electron spectrum shown on Fig.1, every
hole Landau level has splitted into $p$ narrow magnetic subbands grouped near
the unperturbed level. The condition $|V_0|\le\Delta E_{12}$ where
$\Delta E_{12}$ is the distance between levels $n=2+$ and $n=-1-$ allows
to observe the set of non-overlapped magnetic subbands for these levels at
high magnetic fields.

It was mentioned previously that hole Landau levels may be classified into
groupes
of effective Hamiltonian (\ref{htot}) eigenvalues labeled by common index
$n=-1,0,1, \ldots$ For example, for $n=0$ such group belonging to subband
of size quantization with $\nu =1$ consists of one heavy-and one light-hole
level. These levels can be obtained by diagonalization of $2\times 2$ matrix
and are labeled by $n=0-(+)$ (see Fig.4). When the periodic potential of
lateral superlatice is introduced, the $2\times 2$ matrix yields
$2p\times 2p$ matrix equation (\ref{matrix}) which spectrum consists of $2p$
magnetic subbands originating from $n=0-(+)$ levels. If the amplitude $|V_0|$
is small enough to neglect the influence of other levels neighbouring with
the levels $n=0-(+)$, it is possible to study their splitting separately.
The set of $2p$ magnetic subbands originating from levels $n=0-(+)$ splitted
by periodic potential with $V_0=-3 meV$ is shown on Fig.6a(b).
Note that in subbands originating from $n=0-$ level (Fig.6a) the heavy-hole
component with angular momentum $m_J=-3/2$ dominates in the wavefunction while
in subbands splitted from $n=0+$ level (Fig.6b) the light-hole component
with $m_J=-1/2$ has the biggest amplitude.
Comparing Fig.2 and Fig.6, one can see that the difference between electron
and hole spectrum increases at high magnetic fields $q/p \ll 1$ where
the off-diagonal element of Luttinger Hamiltonian
$\gamma_3\sqrt{6eH/c}\ k_z a$ becomes more significant.

In the following Sec. we will calculate the matrix elements for
transitions between valence band and donors located in heterojunction and
thus the knowledge of hole wavefunction in superlattice cell is required.
The real part of hole wavefunction component $m_J=-3/2$ which
dominates among four components of hole wavefunction (\ref{psihole}) in Landau
state $n=-1-$ is shown on Fig.7 at $k_x=k_y=0$. This picture is plotted for
subbands 181 and 200 which are marked by arrows on Fig.5. As for electron
quantum states, in subband 181 located far from the clustering point,
the wavefunction (Fig.7a) has much less zeros then for subband 200
belonging to the region of subbands clustering (Fig. 7b). In detail, the real
part of hole wavefunction on Fig.7a lays below zero almost everywhere and has
a sharp minimum at $x=y=0$. On the opposite, the values of wavefunction for
subband 200 shown on Fig.7b are distributed more uniformly above and below
zero and thus Fig.7b has less dark areas then Fig.7a. We've not shown
the contourplots for imaginary part of wavefunction since they demonstrate
the same behaviour. One can expect that the discussed difference in
wavefunction shape should be reflected in magnitude of matrix elements for
transitions to donors and it will be proved in the following Sec.

When the condition $|V_0|<\Delta E_{12}$ is not fulfilled, the structure of
hole spectrum looks different. The spectrum for $V_0=-10 meV$ and
$\Delta E_{12}\approx 2.5 meV$ is shown on Fig.8. In this case
magnetic subbands originating from different hole Landau levels are strongly
overlapped almost everywhere except the region near the highest Landau level.
This region is marked on Fig.8 and it containes magnetic subbands from 209
to 220 belonging to Landau level $n=2+$. In this interval of
non-overlapping subbands one may expect a distinguishible behaviour of
magnetooptical matrix elements for these subbands (see the following Sec).

Under the conditions of strong subbands overlapp the domination of one of
$|J;m_J\rangle$ basis component becomes less pronounced. This is illustrated
on Fig.9 where all four $|J;m_J\rangle$ components of wavefunction are shown
for subband 185 marked by arrow on Fig.8. It is clearly seen that all
components have the same order which is a consequence of overlapping of those
magnetic subbands originating from Landau levels with different dominating
wavefunction components.

\section{Matrix elements for transitions between conduction (valence) band
and acceptors (donors)}

As it was mentioned in the Introduction, one of possible experimental tools
for investigation of quantum states in magnetic subbands are magnetooptical
measurements of transition intensities. Below we calculate the matrix elements
between Bloch quantum states and impurities located in heterojunction
subjected to magnetic field.

First of all we consider a process in which photon is absorbed and electron
is raised from acceptor atom to electron quantum state (\ref{blochtot})
described in Sec. II. It is supposed that the monolayer of aceptors is
located at well-defined distance from heterojunction interface
\cite{Kukushkin,Volkov}. The initial quantum state $\Psi^I_{x_0y_0}$ is
a wavefunction of shallow acceptor localized at $(x_0,y_0)$ point in $z=z_0$
plane and it has the envelope function of the form
\begin{equation}
\label{accep}
\psi_{x_0y_0}=
A\exp\left(-\frac{1}{r_A}\left[\varrho^2+(z-z_0)^2\right]^{1/2}\right),
\end{equation}

\noindent where $A$ is normalizing constant, $\varrho^2=(x-x_0)^2+(y-y_0)^2$,
$r_A=\kappa_e \hbar^2/m_{val}e^2$ is Bohr radius of acceptor, $\kappa_e$ is
the dielectric constant of material. Here $m_{val}$ stands for averaged
effective mass at the top of valence band with $p$ -\ type atomic function
$p({\bf r})$. We believe that both atomic and envelope functions in
Eq.(\ref{accep}) are practically unaffected by external magnetic field.
The parameters for which the matrtix elements of transitions between acceptors
and conduction band have been calculated were the following: $p/q=20$
(corresponding to $H \approx 12.1 \ T$), the amplitude of periodic potential
$V_0=20 \ meV$ and $a=80 \ nm$. In this case the set of non-overlapping
magnetic subbands has a simple structure shown on Fig.1. For direct optical
transitions one can write \cite{Ancilotto}
\begin{eqnarray}
\nonumber
M_{k_xk_y}(x_0,y_0)=\langle\Psi^F_{k_xk_y}\mid {\bf p\cdot e}
\mid \Psi^I_{x_0y_0}\rangle= \\
\label{mel}
=\langle s\mid {\bf p\cdot e}
\mid p \rangle \langle \psi_{k_x k_y} \mid \psi_{x_0y_0}\rangle+
{\bf e}\cdot \langle \psi_{k_x k_y}\mid {\bf p} \mid \psi_{x_0y_0}\rangle
\langle s\mid p \rangle,
\end{eqnarray}

\noindent where ${\bf e}$ being a unit vector in the direction of incident
electric field and scalar products are defined as
\begin{eqnarray}
\nonumber
\langle s\mid (\ldots) \mid p \rangle=
\int_{cell}s^*({\bf r})(\ldots) p({\bf r})d{\bf r}, \\
\nonumber
\langle \psi_{k_x k_y}\mid(\ldots)\mid \psi_{x_0y_0}\rangle=\int_{crystal}
\psi^*_{k_x k_y}({\bf r})(\ldots)\psi_{x_0y_0}({\bf r})d{\bf r}.
\end{eqnarray}

\noindent The first term in (\ref{mel}) corresponds to matrix elements of
interband transitions from acceptors to conduction band while the second
one has a form of intraband transitions which occur at cyclotron resonance
\cite{DP}. For the problem which is under consideration in this paper
the latter term vanishes due to ortogonality of atomic functions
$p({\bf r})$ and $s({\bf r})$ being $p$- \ and $s$ -\ type functions,
respectively. The uniform distribution of space orientation for $p$ - type
acceptor atomic functions is assumed and one can easily check that due to
this fact the matrix element $\langle s\mid {\bf p\cdot e}\mid p \rangle$
in (\ref{mel}) does not depend on polarization of incident radiation.

In Sec. II it was found that the overlapping of electron and acceptor
wavefunctions and thus the matrix element strongly depend on the position of
acceptor atom in a current superlattice cell. In order to obtain
the transition probability for lateral superlattice with many cells we have
to average it over many possible acceptor positions:
\begin{equation}
{\overline {\mid M \mid ^2_{k_xk_y}}}=\frac{1}{N_A} \sum_{x_0,y_0} \mid
M_{k_xk_y}(x_0,y_0)\mid ^2,
\end{equation}

\noindent where $N_A$ is total number of acceptor positions. It should be
noted that due to the random position of acceptor atom the matrix elements
do not depend on the quasimomentum which classify the Bloch quantum state
(\ref{psiel}). This independance on $k_x$ and $k_y$ reflects
the behaviour of electron wavefunctions which practically remain unchanged
with respect to the variations of quasimomentum. On Fig.10 we plot
the averaged square of matrix element module which determines the transition
probability to the particular subband of Landau levels shown on Fig.1.
In order to compare these values with matrix element for unperturbed Landau
level we plot this matrix element multiplied by $p$ (being the ratio between
the number of states per Landau level and per one magnetic subband) on
the left side of each histogramm of Fig.10 (marked as $LL$). The elements from
1st to 20th (Fig.10a) correspond to transitions to the lowest Landau level
$N=0$, the elements from 21st to 40th (Fig.10b) are plotted for the level
$N=1$ and the elements from 81st to 100th (Fig.10c) describe the transitions
to the level $N=4$. Looking on Fig.10 one can see the huge increase of matrix
elements with respect to subband number inside one Landau level. It reflects
the distribution of electron wavefunction shown on Fig.3: wavefunction in
subbands which are located near clustering point (Fig.3a) have more
oscillations (more zeros) then those related to other edge of splitted Landau
level (Fig.3b). By comparing Fig.10a(b) and Fig.10c one can see the decrease
of matrix elements magnitude both for unperturbed Landau level (marked as
$LL$) and for magnetic subbands in which it has been splitted in. One may
expect that such a decrease is caused by increasing number of wavefunction
oscillations with respect to Landau level index $N$.

The calculation of matrix elements for valence band -- donors transitions
can be investigated similarely to the problem of acceptors -- conduction band
trabsitions discussed above. Namely, the initial quantum state
$\Psi^I_{k_xk_y}$ is now a hole wavefunction (\ref{psihole}), and the final
quantum state $\Psi^F$ is wavefunction of shallow donor impurity located in
the layer $z=z_0$ and described by envelope function $\psi_D({\bf r})$
\begin{eqnarray}
\nonumber
\psi_D=A\exp\left(-\frac{1}{r_D}\left[\varrho^2+(z-z_0)^2\right]^{1/2}\right),
\end{eqnarray}

\noindent where analogous to (\ref{accep}) $r_D=\kappa_e \hbar^2/m_{eff}e^2$
stands for donor Bohr radius (its typical value is $\approx 15 nm$) and
$m_{eff}$ is effective mass at the bottom of conduction band.
This band is characterized by $s$-~type atomic function $s_{\alpha}({\bf r})$
where the index $\alpha=1(2)$ corresponds to the function
$\mid s\uparrow\rangle \left(\mid s\downarrow\rangle \right)$. Since the total
ansamble of donor atoms does not have definite projection of angular momentum,
one can write
\begin{eqnarray}
\nonumber
\Psi^F=\psi_D\frac{\mid s\uparrow\rangle+\mid s\downarrow\rangle}{\sqrt2}.
\end{eqnarray}

\noindent After the definition of initial and final quantum states, we write
the matrix element similiar to (\ref{mel}) as
\begin{eqnarray}
\nonumber
M_{k_xk_y}=\langle\Psi^F \mid {\bf p\cdot e}\mid \Psi^I_{k_xk_y}\rangle= \\
\label{holemel}
=\sum_{\alpha=1}^2\sum_{j=1}^4\langle s_{\alpha}\mid {\bf p\cdot e}
\mid v_j \rangle \langle \psi_D\mid \psi^{(j)}_{k_x k_y}\rangle+
\sum_{\alpha=1}^2\sum_{j=1}^4 {\bf e}\cdot \langle\psi_D\mid{\bf p}
\mid \psi^{(j)}_{k_x k_y}\rangle \langle s_{\alpha}\mid v_j \rangle,
\end{eqnarray}

\noindent where $v_j$ is  $|J;m_J\rangle$ basis function (\ref{jmj}) and
the second term in (\ref{holemel}) again vanishes. On the opposite to
the electronic case, the hole -- donor transition intesities strongly depend
on polarizatoin of incident radiation. On the one hand, it is a consequence
of different contribution of $|J;m_J\rangle$ basis components into hole
quantum state (\ref{psihole}) and, on the other hand, the transitions from
heavy holes are three times more intensive then those from light holes
(see, for example, \cite{Ancilotto,Kukushkin}). The z-dependent calculation
of integral $\langle \psi_D\mid \psi^{(j)}_{k_x k_y}\rangle$ is performed
for functions $c(z)$ for typical GaAs/AlGaAs heterojunction taken
from \cite{Broido}. Similar to the case of transitions between acceptors
and conduction band, we have to average the matrix elements over many possible
donor positions. The averaged squares of matrix elements module
(\ref{holemel}) calculated for two highest Landau levels $n=-1-$ and $n=2+$
being splitted by $V_0=-2.5 meV$ are shown on Fig.11. Here Fig.11a(b)
corresponds to $\sigma^+(\sigma^-)$ polarized radiation. It is evident that
magnetic subbands related to different hole Landau levels exhibit itself
differently. Namely, for $\sigma^+$ polarization the matrix elements for
subbands 181 -- 200 related to $n=-1-$ level are two orders of magnitude
larger then those related to $n=2+$ level. On the contrary, for $\sigma^-$
polarization the elements for subbands 201 -- 220 related to $n=2+$
are five orders of magnitude larger then the elements corresponding
to $n=-1-$ level. We believe that such drastic differences in magnetooptical
parameters will provide more transparence in experimental studies of hole
magnetic subbands. It is obvious that the low amplitude of periodic potential
$|V_0|<\Delta E_{12}$ is important for non-overlap of magnetic subbands which
is illustrated on Fig.12 where matrix elements for the same Landau levels
$n=-1-$ and $n=2+$ splitted by higher periodic potential of quantum dots
$V_0=-10 meV$ are shown. The switching of polarization from $\sigma^+$
(Fig.12a) to $\sigma^-$ (Fig.12b) leads to total decrease of matrix elements
but their internal shape changes significantly mainly for subbands 209 -- 220
which are not overlapped with those related to other Landau levels (see
the marked region on Fig.8). The polarization switching illuminates these
subbands and thus makes possible to detect them experimentally.

\section{Summary and conclusions}

We investigated quantum states and magnetooptics of 2D electrons and holes
in heterojunctions subjected to perpendecular magnetic field and periodic
potential of superlattice. The electron quantum states in {\it n}-type
heterojunction have been studied both for coupled and uncoupled Landau levels
in a wide interval of magnetic field. The holes in {\it p}-type heterojunction
were described by $4\times 4$ Luttinger Hamiltonian where both confinement
potential and potential of lateral surface superlattice have been introduced.
This model allowed us to figure out the influence of spin-orbit coupling onto
four-component Bloch quantum states in external magnetic field. We've
calculated hole magnetic subbands at high magnetic fields under consideration
of several Landau levels originating from the first three subbands of size
quantization. In a wider interval of both low and high magnetic fields
the set of hole magnetic subbands originating from two coupled Landau levels
has been obtained. Here the increasing differences with electron quantum
states occure at high magnetic fields which is caused by the $H$-dependent
off-diagonal term in Luttinger Hamiltonian. Then the calculations of matrix
elements for transitions between electron magnetic subbands and acceptors and
between hole magnetic subbands and donors have been performed. We found
the characteristic dependencies of matrix elements on subband number both
in $n$- and $p$-type heterojunctions. In the latter case the strong dependence
on polarization of incident radiation is found. In particular, at $\sigma^+$
($\sigma^-$) polarization the most intensive transitions are from those hole
magnetic subbands where "spin"-down(up) components of wavefucntion dominate.
The discussed effects allowed us to define the set of parameters (superlattice
periods, amplitude of periodic potential and magnetic field value) for
transparent experimental observation of sharp non-overlapping magnetic
subbands both for electrons and holes.

\section*{Acknowledgments}

We thank A.A. Perov for fruitful discussions and for technical assistance.
This work was supported by the Russian Foundation of Basic Research
(Grant No. 01-02-17102), by the Russian Ministry of Education
(Grant No. E00-3.1-413) and jointly by CRDF Foundation and Russian Ministry
of Education (Project No. REC - 001).


\begin{thebibliography}{99}
\bibitem{Thouless}
D.J. Thouless {\it et al.}, Phys. Rev. Lett. {\bf 49}, 405 (1982).
\bibitem{Silb}
H. Silberbauer, J. Phys.: Condens. Matter {\bf 4}, 7355 (1992).
\bibitem{Geisel}
D. Springsguth, R. Ketzmerick, and T. Geisel, Phys. Rev. B {\bf 56}, 2036
(1997).
\bibitem{DP}
V.Ya. Demikhovskii and A.A. Perov, Phys. Low-Dim. Structures {\bf 7}/{\bf 8},
135 (1998).
\bibitem{Usov}
N. Usov, Sov. Phys. JETP {\bf 67}, 2565 (1988).
\bibitem{Wiegman}
P.B. Wiegman and A.V. Zabrodin, Phys. Rev. Lett. {\bf 72}, 1890 (1994).
\bibitem{Hatsugai}
Y. Hatsugai, M. Kohmoto, and Y.S. Wu, Phys. Rev. Lett. {\bf 73}, 1134 (1994);
Phys. Rev. B {\bf 53}, 9697 (1996).
\bibitem{Krasovsky}
I. Krasovsky, Phys. Rev. B {\bf 59}, 322 (1999).
\bibitem{Eroms}
J. Eroms {\it et al.}, Phys. Rev. B {\bf 59}, R7829 (1999).
\bibitem{Petscel}
G. Petscel and T. Geisel, Phys. Rev. Lett. {\bf 71}, 239 (1993).
\bibitem{Ketzmerick}
R. Ketzmerick {\it et al.}, Phys. Rev. Lett {\bf 84}, 2929 (2000).
\bibitem{Weiss}
D. Weiss {\it et al.}, Phys. Rev. Lett. {\bf 66}, 27 (1991).
\bibitem{Schl}
T. Schl\"osser {\it et al.}, Semicond. Sci. Technol. {\bf 11}, 1582 (1996);
Europhys. Lett. {\bf 33}, 683 (1996).
\bibitem{Albrecht}
C. Albrecht {\it et al.}, Phys. Rev. Lett. {\bf 86}, 147 (2001).
\bibitem{Kukushkin}
I.V. Kukushkin {\it et al.}, Phys. Rev. Lett. {\bf 79}, 1722 (1997).
\bibitem{Volkov}
O.V. Volkov {\it et al.}, Phys. Rev. B {\bf 56}, 7541 (1997).
\bibitem{Luttinger}
J.M. Luttinger, Phys. Rev. {\bf 102}, 1030 (1956).
\bibitem{Broido}
D.A. Broido and L.J. Sham, Phys. Rev. B {\bf 31}, 888 (1985).
\bibitem{Rashba}
Yu.A. Bychkov and E.I. Rashba, in the {\it Proc. of the $17^{th}$
Int. Conf. on the Phys. Semicond.}, San Francisco (1984), Springer Verlag
(1985), p. 321.
\bibitem{Sham}
G.E. Marques and L.J. Sham, Surf.Sci. {\bf 113}, 131 (1982).
\bibitem{Ancilotto}
F. Ancilotto, A. Fasolino, and J.C. Maan, Phys. Rev. B {\bf 38}, 1788 (1988).

\end{thebibliography}
\end{document}